\documentstyle[epsfig]{mn2e}

\def\gtsima{$\; \buildrel > \over \sim \;$}
\def\ltsima{$\; \buildrel < \over \sim \;$}
\def\gtrsim{\lower.5ex\hbox{\gtsima}}
\def\lesssim{\lower.5ex\hbox{\ltsima}}

\begin{document}
%

%A nuclear star cluster surrounding ESO 243-49 HLX-1?
%The nature of ESO 243-49 HLX-1: the gas-rich minor merger scenario
%Is ESO 243-49 HLX-1 the remnant of a gas-rich minor merger?
%The origin of  ESO 243-49 HLX-1 from a gas-rich minor merger
%A minor merger scenario to explain ESO 243-49 HLX-1
\title[A minor merger scenario for HLX-1]{A minor merger scenario for the ultraluminous X-ray source ESO 243-49 HLX-1}
\author[Mapelli et al.]
{M. Mapelli$^{1}$, L. Zampieri$^{1}$, L. Mayer$^{2}$
\\
$^1$INAF-Osservatorio Astronomico di Padova, Vicolo dell'Osservatorio 5, I--35122, Padova, Italy, {\tt michela.mapelli@oapd.inaf.it}\\
$^2$Institute for Theoretical Physics, University of Z\"urich, Winterthurerstrasse 190, CH--8057, Z\"urich, Switzerland\\
}
\maketitle \vspace {7cm }

\begin{abstract}
The point-like X-ray source HLX-1 is the brightest known ultraluminous X-ray source and likely the strongest intermediate-mass black hole candidate. HLX-1 is hosted in the S0 galaxy ESO 243-49, but offset with respect to the nucleus, and its optical counterpart was identified with a massive star cluster. In this paper, we study, through $N-$body/smoothed particle hydrodynamics simulations, the scenario where ESO 243-49 is undergoing (or just underwent) a minor merger with a gas-rich low-mass late-type galaxy. The simulations suggest that the observed star formation rate (SFR) in ESO 243-49 is a consequence of the interaction and that the companion galaxy already underwent the second pericentre passage. We propose that the counterpart of HLX-1 coincides with the nucleus (and possibly with the nuclear star cluster) of the secondary galaxy. We estimate that, if the minor merger scenario is correct, the number density of X-ray sources similar to HLX-1 is $\approx{}10^{-6}$ Mpc$^{-3}$.
%there might be $\approx{}2\times{}10^{-6}$ Mpc$^{-3}$ X-ray sources similar to HLX-1.
\end{abstract}

%
%________________________________________________________________

\section{Introduction}
The point-like X-ray source 2XMM~J011028.1$-$460421 (hereafter, HLX-1), with a maximum luminosity $\sim{}10^{42}$ erg s$^{-1}$ (Farrell et al. 2009, hereafter F09; Godet et al. 2009), is the brightest known ultraluminous X-ray source (ULX, see Feng \&{} Soria 2011 for a review) and likely the strongest intermediate-mass black hole (IMBH) candidate (see van der Marel 2004 for a review). Estimates of the black hole (BH) mass based on the X-ray spectrum range from $\approx{}500$ to $10^4$ M$_\odot{}$ (F09; Davis et al. 2011; Servillat et al. 2011). HLX-1 is located in the outskirts of the S0 galaxy ESO~243-49 (luminosity distance $\sim{}95$ Mpc), $\sim{}0.8$ kpc out of the plane and $\sim{}3.3$ kpc away from the nucleus. HLX-1 has an optical counterpart (Soria et al. 2010, 2012, hereafter S10, S12, respectively; Farrell et al. 2012, hereafter F12), whose association with ESO~243-49 is confirmed by the redshift of the observed H$\alpha{}$ emission line (Wiersema et al. 2010).

{\it Swift}/Ultra-violet optical telescope ({\it Swift}/UVOT) observations of ESO~243-49 show asymmetric ultra-violet (UV) emission at $\sim{}2000$ \AA{} (S10, Webb et al. 2010). The UV emission is centred on the bulge of ESO~243-49, but has an  asymmetric spatial distribution. A background galaxy at  $z\sim{}0.03$ can explain the asymmetry of the UV emission and contributes to a small fraction of %(but cannot account for all) 
the UV emission observed with {\it Swift}/UVOT (F12, S12). The UV emission centred on ESO~243-49 indicates ongoing star formation (SF) at a rate  $\sim{}0.03$ M$_\odot$ yr$^{-1}$, consistent with a younger stellar population superimposed to the $\sim{}5$-Gyr-old dominant stellar component of ESO~243-49 (S10). This  has been interpreted as the fingerprint of a recent minor merger between the S0 galaxy and a gas-rich dwarf galaxy (S10), whose nucleus may be identified with the counterpart of HLX-1. The presence of prominent dust lanes around the nucleus of ESO~243-49 (F12) is another hint of a recent gas-rich merger (e.g., Finkelman et al. 2010; Shabala et al. 2011).  The possibility that minor mergers switch on hyperluminous X-ray sources (HLXs, i.e. ULXs with X-ray luminosity -assumed isotropic- $L_{\rm X}>10^{41}$ erg s$^{-1}$) was initially proposed by King \&{} Dehnen (2005, see also Bellovary et al. 2010).

Furthermore, the optical counterpart of HLX-1 is associated with UV emission  in both the near and the far UV, as shown by {\it Hubble Space Telescope} ({\it HST}) and Very Large Telescope (VLT) photometry (F12; S12). The UV emission from the counterpart of HLX-1 was also found to be variable (S12).

The nature of the HLX-1 counterpart is still debated. Fits to the {\it HST} data (F12) indicate a total stellar mass of $4-6\times{}10^6$ M$_\odot{}$, whereas both a very young ($\sim{}10$ Myr) and a very old ($\sim{}13$ Gyr) age are possible. An intermediate-age solution is not allowed. In the case of a very young stellar population minimal reprocessing from the disc is required, whereas the old population explanation predicts that most of blue/UV emission comes from disc irradiation (F12). On the basis of VLT data, S12 show that the optical/UV component is variable and therefore dominated by disc irradiation, and exclude the scenario of a $\gg{}10^4$ M$_\odot{}$ young star cluster. The two remaining scenarios for the counterpart are a old $\sim{}10^6$ M$_\odot{}$ star cluster (likely a globular cluster) and a $\approx{}10^4$ M$_\odot{}$ young star cluster. The latter explanation is complicated by the fact that such small star clusters unlikely host very massive ($>100$ M$_\odot{}$) BHs (e.g., Portegies Zwart \&{} McMillan 2002).

 A further issue is represented by the  observed X-ray variability of HLX-1, with a semi-regular period of $\sim{}380$ days (Servillat et al. 2011), and by the mechanism driving the accretion. Lasota et al. (2011, hereafter L11) propose that the  observed X-ray variability may reflect the orbital period of the companion star. In this case, the outbursts are triggered by periastron passages for a sufficiently eccentric orbit of the companion star (so that the tidal disruption radius is of the order of the periapsis distance). Considering the high mass transfer rate required to power HLX-1 ($\approx{}10^{-4}$ M$_\odot{}$ yr$^{-1}$), L11 suggest that the companion star of the IMBH can be an asymptotic giant branch (AGB) star and that its mass cannot be too small ($<<3$ M$_\odot{}$).
%thermal-viscous instabilities in the accretion disc are unlikely to produce the  observed X-ray variability of HLX-1. They argue that the observed X-ray variability may reflect the orbital period of the companion star and that the outbursts are triggered by periastron passages for a sufficiently eccentric orbit (so that the tidal disruption radius is of the order of the periapsis distance). In this scenario, considering the high mass loss required by HLX-1 properties ($\approx{}10^{-4}$ M$_\odot{}$ yr$^{-1}$), L11 suggest that the companion star of the IMBH can be an asymptotic giant branch (AGB) star and that its mass cannot be too small ($<<3$ M$_\odot{}$).}
%. The mass of the AGB star cannot be too small ($<<3$ M$_\odot{}$), not to be destroyed too early by the IMBH. 

%comes from the fact that the mass of the companion star of the IMBH cannot be too small ($<<3$ M$_\odot{}$) to reproduce the observed X-ray variability (see L11). This requires the association of HLX-1 with a relatively young stellar population. 

In this paper, we propose a new interpretation connected with the minor merger scenario (initially proposed by S10). We investigate, through $N-$body/smoothed particle hydrodynamics (SPH) simulations, the possibility that the optical counterpart of HLX-1 is the nucleus of a gas-rich low-mass late-type galaxy, undergoing merger with ESO 243-49. In line with this interpretation, we suggest that HLX-1 is associated with the central BH of the disrupted galaxy and with its surrounding nuclear star cluster (NC). This entails that the counterpart of HLX-1 consists mainly of very old ($> 10$ Gyr) stars, but has also a younger population. In fact, NCs are known to have multiple stellar populations (Rossa et al. 2006; Walcher et al. 2006), generally dominated by a old component, but where a fraction of young stars is also present (e.g., the case of the Milky Way, see Pfuhl et al. 2011). Our simulations show that the overall scenario in which HLX-1 is the nucleus of a stripped, gas-rich low-mass galaxy is consistent with the available observations.
% We suggest that HLX-1 is associated with the central BH of the dwarf and with its surrounding NC. In fact, NCs are known to have multiple stellar populations (Rossa et al. 2006; Walcher et al. 2006), generally dominated by a old component, but were a fraction of young stars is also present (e.g., the case of the Milky Way, see Pfuhl et al. 2011). 
\subsection{The minor merger scenario}
A minor merger scenario between the S0 and a gas-rich low-mass galaxy would naturally explain the UV emission and the dust lanes observed in the central region of ESO 243-49, as well as (at least) part of the UV emission in coincidence with the HLX-1 counterpart. Furthermore, a minor merger scenario has other intriguing implications. For example, the counterpart of HLX-1 can be identified with the nucleus of the disrupting galaxy. This provides a possible explanation for the mass of the BH in HLX-1, as the nuclei of dwarf galaxies have long been suspected to harbour the low-mass tail of super-massive BHs (SMBHs, see, e.g., Barth et al. 2004; Peterson et al. 2005; van Wassenhove et al. 2010; Reines et al. 2011). Recent studies (Graham 2012a, 2012b) indicate that, in low-mass galaxies, the mass of the central BH scales approximately with the square of the host spheroid mass (rather than linearly, as was usually thought, e.g. Marconi \&{} Hunt 2003), implying that low-mass galaxies with a $10^8-10^9$ M$_\odot{}$ bulge can host BHs with a mass $\approx{}10^3-10^5$ M$_\odot{}$, consistent with the expected mass of the BH in HLX-1. Finally, nuclei of galaxies in this mass range frequently host NCs (e.g., B\"oker et al. 2002; Graham \&{} Guzm\'an 2003; C$\hat{\rm o}$t\'e et al. 2006; Graham \&{} Spitler 2009; see B\"oker 2010 for a recent review). The presence of a central BH and that of a NC do not seem to be mutually exclusive: in addition to the Milky Way (e.g., Sch\"odel, Merritt \&{} Eckart 2009 and references there), Filippenko \&{} Ho (2003) identified at least one galaxy (NGC~4395) hosting both a SMBH and a NC, and Graham \&{} Driver (2007) subsequently reported the existence of two additional such galaxies (NGC~3384 and NGC~7457). The sample of galaxies hosting both a SMBH and a NC was substantially increased by  Gonzalez Delgado et al. (2008, 2009), Seth et al. (2008) and Graham \&{} Spitler (2009). Unlike most of the Galactic globular clusters, NCs have often a complex SF history, with multiple episodes of SF (Rossa et al. 2006; Walcher et al. 2006). This suggests the possibility that the colours of the HLX-1 counterpart are the result of  multiple populations, dominated by old stars, but with a smaller contribution by very young stars (plus possibly emission from X-ray reprocessing in the disc).

We therefore simulate different scenarios for the merger between a S0 galaxy and a gas-rich companion, to check whether the simulated kinematics and SF history match the observed properties of ESO 243-49 and of the counterpart of HLX-1. 
%In addition, the spheroid and the mass of the BH is steeper than linear   
\section{Method: $N-$body simulations}
The initial conditions for both the primary galaxy (i.e. the model of ESO 243-49) and the secondary galaxy are generated by using an upgraded version of the code described in Widrow, Pym \&{} Dubinski (2008; see also Kuijken \&{} Dubinski 1995 and Widrow \&{} Dubinski 2005). The code generates self-consistent disc-bulge-halo galaxy models, derived from explicit distribution functions for each component, that are very close to equilibrium. In particular, the halo is modelled as a Navarro, Frenk \&{} White (1996, NFW) profile. We use an exponential disc model (Hernquist 1993), while the bulge is spherical and comes from a generalization of the Sersic law (Prugniel \&{} Simien 1997; Widrow et al. 2008).
 
Both the primary and the secondary galaxy have a stellar bulge and a stellar disc.
 The giant S0 galaxy has no gas, whereas the secondary galaxy has an initial gas mass of $1.38\times{}10^{8}$ M$_\odot{}$, distributed according to an exponential disc. Therefore, the initial configuration of the secondary galaxy is consistent with a low-mass gas-rich disc galaxy. The total mass of the secondary is $\sim{}1/20$ of the mass of the primary, classifying the outcome of the interaction as a minor merger. The masses of the various components and the characteristics lengths of the simulated galaxies are listed in Table~1.

%For runs A, B, C, D, F and H, the impact parameter is $b=8$ kpc. In runs E and G $b$ is $=12$ kpc. 
Here, we report the results of two different runs, in which 
%the simulated galaxies are identical 
the masses and scale lengths of each galaxy are the same (as described in Table~1), but their orbital properties (impact parameter, relative velocity, orientation angles and total energy, listed in Table~2) are different.
The main common orbital feature between the two runs is that the centre of mass (CM) of the secondary galaxy is 
%disc of the two galaxies are 
assumed to lie approximately on the same plane  as the disc of the primary galaxy (in run~A they initially lie exactly on the same plane, whereas in run~B there is an initial shift of 2 kpc). This was required as the offset between the HLX-1 counterpart and the disc of the S0 galaxy is relatively small ($\sim{}0.8$ kpc).
%, and the velocities are approximately the same (the counterpart of HLX-1 being `red-shifted' by about 170 km s${^-1}$, Wiersema et al. 2010). 
We initially set the CM of the secondary galaxy at a distance of $D=200$ kpc and $150$ kpc from the CM of the primary in runs~A and B, respectively (Table~2). These distances are larger than two virial radii of the primary.
The adopted orbits are nearly parabolic (run~A is slightly hyperbolic and run~B is slightly bound) and with very high eccentricity (Table~2), in agreement with predictions from cosmological simulations (Khochfar \&{} Burkert 2006).

The particle mass in the primary galaxy is $2.5\times{}10^5$ M$_\odot{}$ and $5\times{}10^4$ M$_\odot{}$ for dark matter (DM) and stars, respectively. The particle mass in the secondary galaxy is  $2.5\times{}10^4$ M$_\odot{}$ for DM and $5\times{}10^3$ M$_\odot{}$ for both stars and gas\footnote{We checked by running simulations with different resolution that our choice of particle masses does not affect significantly the dynamics of the system (e.g., by inducing appreciable spurious dynamical friction or ejections). We do not find significant differences, apart from a change in the resolution of the SF.}. 
The softening length is 0.1 kpc. We integrate the systems for 4 Gyr after the first pericentre passage.
We simulate the evolution of the models with the $N-$body/SPH tree code gasoline (Wadsley, Quinn \&{} Stadel 2004). Radiative cooling, SF
 and supernova (SN) blastwave feedback are enabled, as described in Stinson et al. (2006, 2009, see also Katz 1992). The adopted parameters for SF and feedback  are the same as used in recent cosmological simulations capable of forming realistic galaxies in a wide range of masses (e.g., Governato et al. 2010; Guedes et al. 2011), and in recent simulations of galaxy-galaxy collisions (Mapelli \&{} Mayer 2012).
 %%%%%%%%%%%%%%%%%%%%%%%%%%%%%%% TABLE 1%%%%%%%%%%%%%%%%%%%%%%%%%%%%%%%%%
\begin{table}
\begin{center}
\caption{Initial conditions: masses and scale lengths.}
 \leavevmode
\begin{tabular}[!h]{lll}
\hline
Model properties & Primary & Secondary \\
\hline
DM Mass [$10^{11}$ M$_\odot{}$]          & 7.0  & 0.3\\
$M_\ast{}$$^{\rm a}$ [$10^{10}$ M$_\odot{}$]      & 7.0   & 0.2\\
$f_{\rm b/d}$                               &  0.25 & 0.25 \\
Gas Mass$^{\rm b}$ [$10^{8}$ M$_\odot{}$]    & 0 &  1.38 \\
Halo scale length$^{\rm c}$ [kpc] & 6.0 & 3.0\\
Disc scale length [kpc] & 3.7 & 3.0 \\
Disc scale height [kpc] & 0.37 & 0.30 \\
Bulge scale length [kpc] & 0.6 & 0.6 \\
\noalign{\vspace{0.1cm}}
\hline
\end{tabular}
\footnotesize{\\$^{\rm a}$  $M_\ast{}$ is the total stellar mass of the galaxy (including both bulge and disc). $f_{\rm b/d}$  is the bulge-to-disc mass ratio. \\$^{\rm b}$The primary has no gas, while the gas of the secondary is distributed according to an exponential disc, with the same parameters (scale length and height) as the stellar disc. $^{\rm c}$ We name halo scale length the NFW scale radius $R_{\rm s}\equiv{}R_{200}/c$, where $R_{200}$ is the virial radius of the halo (NFW 1996) and $c$ the concentration (here we assume $c=12$ for both galaxies).}
%$^{\rm c}$The rate $R$ is given for four different values of the MF slope $\alpha{}=2.35,1.5,1,0,-1$.}
\end{center}
\end{table}
%%%%%%%%%%%%%%%%%%%%%%%%%%%%%%%%%%%%%%%%%%%%%%%%%%%%%%%%%%%%%%%%%%%%%%%%%%%%%

 %%%%%%%%%%%%%%%%%%%%%%%%%%%%%%% TABLE 2%%%%%%%%%%%%%%%%%%%%%%%%%%%%%%%%%%%%%
\begin{table}
\begin{center}
\caption{Initial conditions: orbital parameters.}
 \leavevmode
\begin{tabular}[!h]{lll}
\hline
Orbital parameters & Run A
 & Run B\\
\hline
Impact Parameter $b$ [kpc]         & 10.0   & 10.2      \\
Relative velocity $v_{\rm rel}$ [km s$^{-1}$]         & 200   & 100     \\
$\theta{}$, $\phi{}$, $\psi{}$ [rad]$^{\rm a}$  & $\pi{}/2$, $\pi{}$, 0 & $\pi{}/2$, 0, 2.94   \\
Initial distance $D$ [kpc] & 200 & 150 \\
$E_{\rm s}$ [10$^4$ km$^2$ s$^{-2}$] $^{\rm b}$ & 0.38 & -1.65 \\%0.23 & -0.78 \\
$L_{\rm s}$ [10$^3$ km s$^{-1}$ kpc] $^{\rm c}$ & 2.0 & 1.0 \\
$e$  $^{\rm d}$ & 1.003 & 0.997 \\
Orbit $^{\rm e}$ & prograde & retrograde  \\
\noalign{\vspace{0.1cm}}
\hline
\end{tabular}
\footnotesize{\\$^{\rm a}$ For the definition of $\theta{}$, $\phi{}$, $\psi{}$, see figure~1 of Hut \&{} Bahcall (1983). In particular, $\theta{}$ is the angle between the relative velocity vector ${\bf v}_{\rm rel}$ and the symmetry axis of the primary disc, $\phi{}$ describes the orientation of  ${\bf v}_{\rm rel}$ projected in the plane of the primary disc and $\psi{}$ describes the orientation of the initial distance vector ${\bf D}$ (between the CMs of the two galaxies) in the plane perpendicular to ${\bf v}_{\rm rel}$.\\
%of ${\bf v}_{\rm rel}$ in the plane perpendicular to the initial distance vector. \\
$^{\rm b}$ $E_{\rm s}$ is the  specific orbital energy , i.e. the total energy divided by the reduced mass $\mu{}=m_1\,{}m_2/(m_1+m_2)$ (where $m_1$ and $m_2$ are the mass of the primary and of the secondary galaxy, respectively). $E_{\rm s}\equiv{}-G\,{}M/D+\,{}v_{\rm rel}^2/2$, where $M=m_1+m_2$ is the total mass of the two galaxies, 
%($m_1$ and $m_2$ are the mass of the primary and of the secondary galaxy, respectively),
%  $\mu{}=m_1\,{}m_2/M$ is the reduced mass, 
$G$ is the gravitational constant and $D$ the initial distance between the CMs. \\$^{\rm c}$ $L_{\rm s}$ is the modulus of the specific orbital angular momentum, i.e. the angular momentum divided by the reduced mass. \\$^{\rm d}$ $e$ is the eccentricity ($e=[1+2\,{}E_{\rm s}\,{}L_{\rm s}^2/(G\,{}M)^2]^{1/2}$). \\$^{\rm e}$ A orbit is classified as prograde/retrograde depending on the alignment/counter-alignment of the orbital angular momentum of the secondary galaxy with respect to the spin of the primary galaxy.}
%$^{\rm c}$The rate $R$ is given for four different values of the MF slope $\alpha{}=2.35,1.5,1,0,-1$.}
\end{center}
\end{table}
%%%%%%%%%%%%%%%%%%%%%%%%%%%%%%%%%%%%%%%%%%%%%%%%%%%%%%%%%%%%%%%%%%%%%%%%%%%%%

\section{Results and discussion}
We followed the interaction between the primary (S0) and secondary (gas-rich) galaxy for 4 Gyr after the first pericentre passage (corresponding to $\sim{}5$ Gyr starting from the aforementioned initial conditions). The simulations indicate that the merger phase can be very long-lived ($>3$ Gyr), if the orbital angular momentum (orbital velocity) of the secondary is low (high), as the secondary spends most of its time out of the disc of the primary. We note that a system like ESO 243-49 plus the HLX-1 optical counterpart is not necessarily the result of the last stages of the merger, but can form almost anytime after the first pericentre passage\footnote{The request that the two galaxies already had a first pericentre passage is necessary to justify the estimated SFR from the centre of ESO 243-49 (see Table~3).}, because of projection effects. The only limitation from the available spectroscopic measurements is that the relative velocity between the CMs of the two galaxies is not too large, as the measurement of the H$\alpha{}$ line shows that the HLX-1 counterpart may be `red-shifted' by $\approx{}170$ km s$^{-1}$ with respect to the centre of ESO 243-49 (however, this measurement is quite uncertain, see Wiersema et al. 2010).

\begin{figure}
\center{{
\epsfig{figure=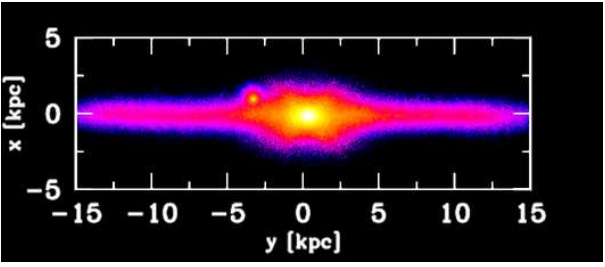,width=8cm} 
}}
\caption{\label{fig:fig1}
%Projected mass density of stars in run~B at $t=0.1$ Gyr after the first pericentre passage. The two galaxies have been projected to the plane that best matches the observed projected position of HLX-1 counterpart with respect to ESO 243-49. The scale is logarithmic, ranging from 2.23 M$_\odot{}$ pc$^{-2}$ to $2.23\times{}10^4$ M$_\odot{}$ pc$^{-2}$.
Projected mass density of stars in run~A at $t=2.6$ Gyr after the first pericentre passage. The two galaxies have been projected to the plane that best matches the observed projected position of the HLX-1 counterpart with respect to ESO 243-49. The scale is logarithmic, ranging from 2.23 M$_\odot{}$ pc$^{-2}$ to $2.23\times{}10^4$ M$_\odot{}$ pc$^{-2}$. The line-of-sight velocity of the secondary galaxy is `red-shifted'  by $\sim{}230$ km s$^{-1}$ with respect to the CM of the primary. 
}
\end{figure}

\begin{figure}
\center{{
\epsfig{figure=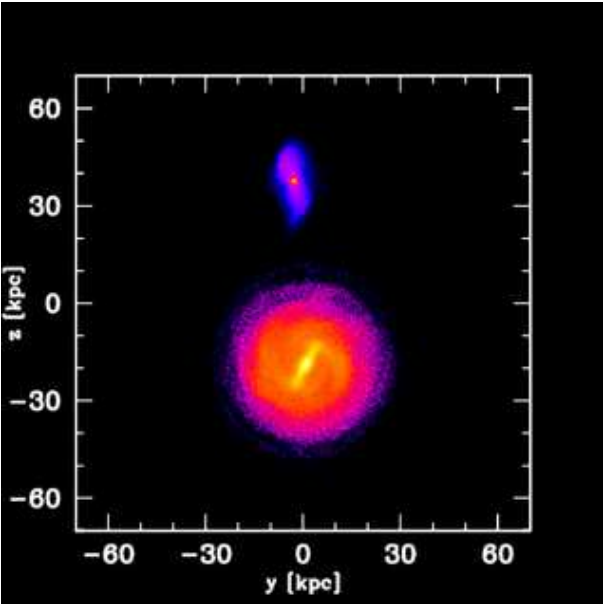,width=8cm} 
}}
\caption{\label{fig:fig2}
%Projected mass density of stars in run~B at $t=0.1$ Gyr after the first pericentre passage. The two galaxies have been projected to the plane that best matches the observed projected position of HLX-1 counterpart with respect to ESO 243-49. The scale is logarithmic, ranging from 2.23 M$_\odot{}$ pc$^{-2}$ to $2.23\times{}10^4$ M$_\odot{}$ pc$^{-2}$.
Projected mass density of stars in run~A at $t=2.6$ Gyr after the first pericentre passage. The two galaxies have been projected so that the primary is seen face-on. The scale is logarithmic, ranging from  $2.23\times{}10^{-2}$ M$_\odot{}$ pc$^{-2}$ to $2.23\times{}10^4$ M$_\odot{}$ pc$^{-2}$. 
}
\end{figure}

This fact is evident from Figs.~\ref{fig:fig1} and \ref{fig:fig2}, that show the projected mass density of the overall stellar component of the two galaxies in run~A (similar considerations can be done for run~B). In  Fig.~\ref{fig:fig1}, the density of stars has been projected in the plane that best matches the observed position of  the HLX-1 counterpart with respect to ESO 243-49: the centre of the secondary is $\sim{}0.8$ kpc out of the plane defined by the disc of the S0 and $\sim{}3.3$ kpc far from the nucleus of the primary. If we measure the relative line-of-sight velocity ($v_{\rm LOS}$) between the CM of  the secondary and of the primary galaxy according to this projection, we find $v_{\rm LOS}=230$ km s$^{-1}$ with respect to the CM of the primary, where the plus sign indicates that the secondary is receding from the observer. 

%%%%%%%%%%%%%%%%%%%%%%%%%%%%%%%%%%% FIGURE 1 %%%%%%%%%%%%%%%%%%%%%%%%%%%%%%%%%%
%%%%%%%%%%%%%%%%%%%%%%%%%%%%%%%%%%%%%%%%%%%%%%%%%%%%%%%%%%%%%%%%%%%%%%%%%%%%%%%
%%%%%%%%%%%%%%%%%%%%%%%%%%%%%%%%%%% FIGURE 2 %%%%%%%%%%%%%%%%%%%%%%%%%%%%%%%%%%
%%%%%%%%%%%%%%%%%%%%%%%%%%%%%%%%%%%%%%%%%%%%%%%%%%%%%%%%%%%%%%%%%%%%%%%%%%%%%%%

 %%%%%%%%%%%%%%%%%%%%%%%%%%%%%%% TABLE 3%%%%%%%%%%%%%%%%%%%%%%%%%%%%%%%%%
\begin{table}
\begin{center}
\caption{SFR and orbital properties in the simulations.}
 \leavevmode
\begin{tabular}[!h]{llllll}
\hline
 & {\it t}
 & SFR 
 & SFR$_{\rm sec}$ 
 & $d_3$ 
 & $v_{\rm LOS}$ \\
 & [Gyr]
 & $[10^{-3}{\rm M}_\odot{}\,{}{\rm yr}^{-1}]$ 
 & $[10^{-3}{\rm M}_\odot{}\,{}{\rm yr}^{-1}]$
 & [kpc] 
 & $[{\rm km}\,{}{\rm s}^{-1}]$  \\
\hline
Run~A  & 0.3 & 1.0  &  1.0   &  89  & 184 \\
       & 0.5  & 2.2  & 2.2    & 118  & 128 \\
       & 1.0  & 1.9  & 0.5    & 154  & 43 \\
       & 1.5  & 4.5  & 0.4    & 152  & -39 \\
       & 2.0  & 20.0 & 0.1    & 110  & -128 \\
       & 2.5  & 48.5 & 0.2     &  29  & 306 \\
       & 3.0  & 22.1 & $<0.1$  & 112  & 73 \\
       & 3.5  & 12.0 & $<0.1$  & 120  & -38 \\
       & 4.0  & 10.4 &  0.0  &  62  & -213 \\
\noalign{\vspace{0.3cm}} 
Run~B  & 0.3  & 1.1  & 1.1     &   66 & 71   \\
       & 0.5  & 0.9  & 0.8     &   65 & -53  \\
       & 1.0  & 1.3  & $<0.1$  &   54 & 140   \\
       & 1.5  & 4.7  & $<0.1$  &   26 & -278  \\
       & 2.0  & 8.1  & 0.0     &   71 & 14 \\
       & 2.5  & 15.6 & 0.0     &   34 & 296 \\
       & 3.0  & 12.9 & 0.0     &   87 & -12 \\
       & 3.5  &  6.6 & 0.0     &    7 & -370 \\
       & 4.0  &  5.2 & 0.0     &   72 & -34 \\
\noalign{\vspace{0.1cm}}
\hline
\end{tabular}
\footnotesize{\\{\it t}: elapsed time since the first pericentre passage; SFR: total SFR; SFR$_{\rm sec}$: SFR in the secondary; $d_3$: three-dimensional distance between the CMs of the two galaxies; $v_{\rm LOS}$: relative velocity between the CMs of the two galaxies along the line of sight. A different line of sight is chosen for each snapshot, to match the observed position of the HLX-1 counterpart with respect to ESO 243-49.} 
\end{center}
\end{table}
%%%%%%%%%%%%%%%%%%%%%%%%%%%%%%%%%%%%%%%%%%%%%%%%%%%%%%%%%%%%%%%%%%%%%%%%%%%%%
Fig.~\ref{fig:fig2} shows that the apparent coincidence of the secondary with the position of the primary when projected as in Fig.~\ref{fig:fig1} is partially  an effect of projection, as the centres of the two galaxies are $\sim{}60$ kpc far from each other. Fig.~\ref{fig:fig2} also shows the large bar ($\sim{}13$ kpc) in the primary, whose formation was induced by the interaction and which produces the boxy shape of the bulge\footnote{A number of (currently unavailable) kinematic data (e.g., a rotation curve) are necessary to constrain the existence of a bar in ESO~243-49 (e.g., Bureau \&{} Athanassoula 1999; Athanassoula \&{} Bureau 1999; Bureau \&{} Athanassoula 2005).}.
%\footnote{The existence of a bar in ESO~243-49 might be inferred by the prominent dust lanes (e.g., Athanassoula 1992), evident in the {\it HST} images (F12).}.
Finally, the tidal tails surrounding the secondary are also apparent from Fig.~\ref{fig:fig2}. Figs.~\ref{fig:fig1} and \ref{fig:fig2} represent the state of the system at $t=2.6$ Gyr after the first pericentre passage (and the first pericentre passage occurred $\sim{}700$ Myr after the beginning of the simulation). At this time, the secondary already passed twice through the inner disc of the primary and it is receding again towards the apocentre.

The last two columns of Table~3 show the three-dimensional ($d_3$) distance and the relative line-of-sight velocity ($v_{\rm LOS}$) between the CMs of the two galaxies at different times $t$, assuming (for each snapshot) the projection that best matches the observed location of the counterpart of HLX-1. As defined before, the plus (minus) sign indicates that the secondary is receding (approaching) with respect to the observer (assuming that the primary galaxy is at rest with respect to the observer), i.e. it is red-shifted (blue-shifted). The modulus of $v_{\rm LOS}$ is almost always consistent with the observations (which suggest a red-shifting by $\approx{}170$ km s$^{-1}$ with a large uncertainty, Wiersema et al. 2010), as a consequence of the assumed initial relative velocity between the two galaxies (Table~2). Therefore, the simulated systems spend most of their time in a kinematic state consistent with the available observations of HLX-1: more accurate spectroscopic measurements of the HLX-1 counterpart are required to put stronger constraints.

%%%%%%%%%%%%%%%%%%%%%%%%%%%%%%%%%%% FIGURE 3 %%%%%%%%%%%%%%%%%%%%%%%%%%%%%%%%%%
\begin{figure}
\center{{
\epsfig{figure=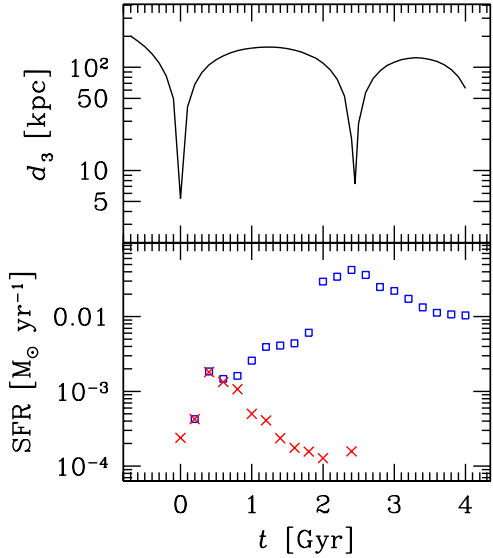,width=8cm} 
}}
\caption{\label{fig:fig3}
Run~A, top panel: three dimensional distance between the CMs of the two galaxies ($d_3$) as a function of time; bottom panel: SFR as a function of time for the entire simulation (open squares, blue on the web) and for the secondary galaxy (crosses, red on the web). $t=0$ is the time of the first pericentre passage.
}
\end{figure}
%%%%%%%%%%%%%%%%%%%%%%%%%%%%%%%%%%%%%%%%%%%%%%%%%%%%%%%%%%%%%%%%%%%%%%%%%%%%%%%

%%%%%%%%%%%%%%%%%%%%%%%%%%%%%%%%%%% FIGURE 4 %%%%%%%%%%%%%%%%%%%%%%%%%%%%%%%%%%
\begin{figure}
\center{{
\epsfig{figure=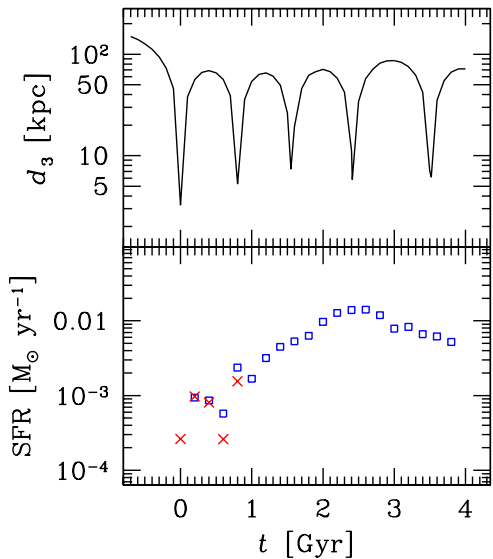,width=8cm} 
}}
\caption{\label{fig:fig4}
The same as Fig.~\ref{fig:fig3}, but for run~B.
}
\end{figure}
%%%%%%%%%%%%%%%%%%%%%%%%%%%%%%%%%%%%%%%%%%%%%%%%%%%%%%%%%%%%%%%%%%%%%%%%%%%%%%%
\subsection{The star formation rate}
Further hints about the formation and evolution of the ESO~243-49 system can be derived from the study of the simulated SF rate (SFR). We remind that the UV observations indicate a SFR$\sim{}0.03$ M$_{\odot{}}$ yr$^{-1}$ centred in the bulge of the S0, although with an asymmetric distribution pointing towards the location of the HLX-1 counterpart (S10). UV emission is also associated with the counterpart of HLX-1 (F12; S12), but it may be totally or partially due to the reprocessing of the HLX-1 accretion disc (S12). 

The SF history from our simulations is shown in Figs.~\ref{fig:fig3} and \ref{fig:fig4}, as well as in Table~3. In particular, the bottom panels of Figs.~\ref{fig:fig3} and \ref{fig:fig4} show the SFR in runs A and B, respectively. The top panels of Figs.~\ref{fig:fig3} and \ref{fig:fig4} show $d_3$ as a function of time in runs A and B, respectively. It is important to compare the distance between the  two galaxies with the SFR, to have an idea of the influence of the pericentre passages on the SF.
In both runs~A and B, there is no SF before the first pericentre passage (that is the absence of points at $t<0$ in Figs.~\ref{fig:fig3} and \ref{fig:fig4} is 
%not due to absence of snapshots, but to the fact that zero gas particles were converted to stars at $t<0$)
due to the fact that no gas particles were converted to stars at $t<0$). This indicates the robustness of our numerical SF recipes (see, e.g., Stinson et al. 2006) and the stability of our initial conditions, as no spurious SF is induced in the first stages of the simulations. SF starts a few Myr after the first pericentre passage. Initially, most of the global SF is concentrated in the gas-rich nucleus of the secondary. 

\begin{figure}
\center{{
\epsfig{figure=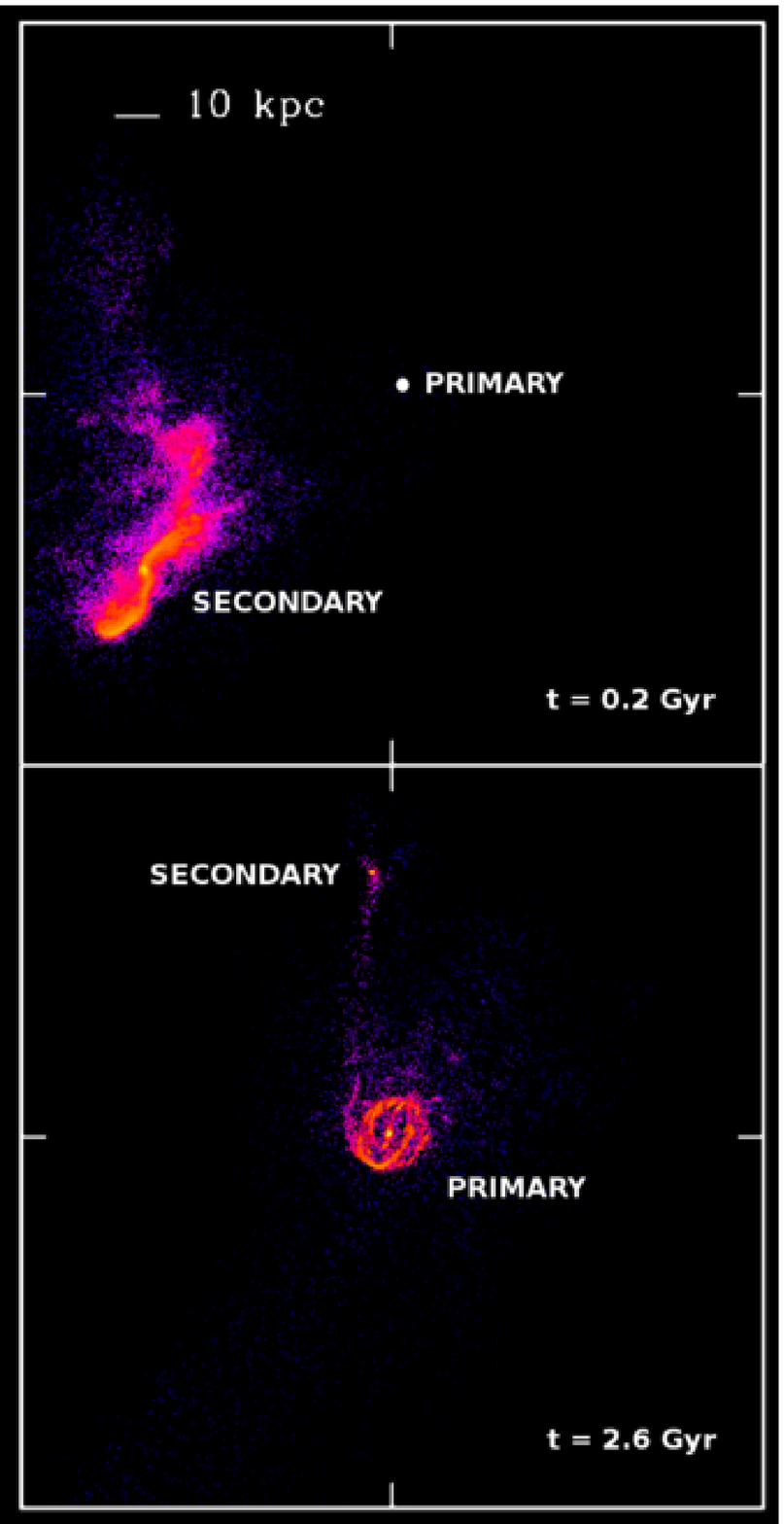,width=7.0cm} 
}}
\caption{\label{fig:fig5}
%Projected mass density of stars in run~B at $t=0.1$ Gyr after the first pericentre passage. The two galaxies have been projected to the plane that best matches the observed projected position of HLX-1 counterpart with respect to ESO 243-49. The scale is logarithmic, ranging from 2.23 M$_\odot{}$ pc$^{-2}$ to $2.23\times{}10^4$ M$_\odot{}$ pc$^{-2}$.
Projected mass density of gas in run~A at $t=0.2$ Gyr (top panel) and at $t=2.6$ Gyr (bottom panel) after the first pericentre passage. The two galaxies have been projected so that the primary is seen face-on. The CM of the primary coincides with the centre of the frames, and it is marked by a white circle in the top panel. Each frame is 160 kpc per edge.
The scale is logarithmic, ranging from $2.23\times{}10^{-6}$ M$_\odot{}$ pc$^{-2}$ to $2.23\times{}10^3$ M$_\odot{}$ pc$^{-2}$. 
}
\end{figure}

After $\sim{}1-2$ Gyr (but this value depends on the orbital properties), most of gas is tidally stripped from the secondary galaxy and accreted by the primary (see Fig.~\ref{fig:fig5}). Given the low angular momentum of the selected orbits, the gas flows almost radially to the centre of the primary, where it starts corotating with the bar, forming a ring at a radius $\approx{}5-8$ kpc, and is partially funnelled towards the centre of the S0. Here, a burst of SF takes place, delayed by $\sim{}1-2$ Gyr with respect to the burst in the nucleus of the secondary. The SF burst in the centre of the primary is  particularly strong after the second (or more) pericentre passage of the secondary.

%%%%%%%%%%%%%%%%%%%%%%%%%%%%%%%%%%% FIGURE 5 %%%%%%%%%%%%%%%%%%%%%%%%%%%%%%%%%%
%%%%%%%%%%%%%%%%%%%%%%%%%%%%%%%%%%%%%%%%%%%%%%%%%%%%%%%%%%%%%%%%%%%%%%%%%%%%%%%

 %%%%%%%%%%%%%%%%%%%%%%%%%%%%%%% TABLE 4%%%%%%%%%%%%%%%%%%%%%%%%%%%%%%%%%
\begin{table*}
\begin{center}
\caption{Evolution of the stellar and gas mass bound to the secondary galaxy.}
 \leavevmode
\begin{tabular}[!h]{llllllll}
\hline
 & {\it t}
 & $M^{\rm sec}_{\rm g,\,{}0.35}$
 & $M^{\rm sec}_{\rm g,\,{}1}$ 
 & $M^{\rm sec}_{\rm \ast{},\,{}0.35}$
 & $M^{\rm sec}_{\rm \ast{},\,{}1}$ 
 & $M^{\rm sec}_{\rm y,\,{}0.35}$
 & $M^{\rm sec}_{\rm y,\,{}1}$\\
 & [Gyr]
 & [$10^6$ M$_\odot{}$]
 & [$10^6$ M$_\odot{}$]
 & [$10^7$ M$_\odot{}$]
 & [$10^7$ M$_\odot{}$]
 & [$10^5$ M$_\odot{}$]
 & [$10^5$ M$_\odot{}$]\\
\hline
Run  A & 0.3 & 4    & 13   & 15  & 35  & 1.1    & 1.1   \\
       & 0.5 &  4   & 12   & 14  & 34  & 3.7    & 4.7   \\
       & 1.0 &  2   & 4    & 14  & 33  & 1.9    & 1.9   \\
       & 1.5 &  2   & 4    & 14  & 33  & 0.6    & 0.6   \\
       & 2.0 &  2   & 2.5  & 14  & 33  & 0.4    & 0.4   \\
       & 2.5 &  0.8 & 1.4  & 9.1 & 22  & 0.4    & 0.4   \\
       & 3.0 &  1.2 & 1.4  & 9.1 & 20  & 0.1    & 0.1   \\
       & 3.5 &  0.8 & 0.9  & 8.8 & 20  & 0.1    & $<0.1$\\
       & 4.0 &  0.8 & 0.9  & 9.0 & 19  & $<0.1$ & $<0.1$\\
\noalign{\vspace{0.3cm}} 
Run B  & 0.3 & 4    & 12   & 12  & 25  & 2.3    & 2.5    \\
       & 0.5 & 3.4  & 11   & 12  & 25  & 1.9    & 2.0    \\ 
       & 1.0 &  0.9 & 2    & 7.2 & 13  & 1.5    & 2.8    \\
       & 1.5 &  1.4 & 2    & 7.1 & 13  & $<0.1$ & $<0.1$ \\
       & 2.0 & 0.1  & 0.1  & 3.9 & 6.5 & 0.0    & 0.0    \\
       & 2.5 & 0.0  & 0.0  & 1.4 & 3.2 & 0.0    & 0.0    \\
       & 3.0 & 0.0  & 0.0  & 1.4 & 2.3 & 0.0    & 0.0    \\
       & 3.5 & 0.0  & 0.0  & 1.8 & 1.8 & 0.0    & 0.0    \\
       & 4.0 & 0.0  & 0.0  & 0.4 & 0.6 & 0.0    & 0.0    \\
\noalign{\vspace{0.1cm}}
\hline
\end{tabular}
\footnotesize{\\{\it t}: elapsed time since the first pericentre passage;
$M^{\rm sec}_{\rm g,\,{}0.35}$ and $M^{\rm sec}_{\rm g,\,{}1}$: total mass of gas in the inner 0.35 kpc and 1 kpc of the secondary galaxy, respectively; $M^{\rm sec}_{\rm \ast{},\,{}0.35}$ and  $M^{\rm sec}_{\rm \ast{},\,{}1}$: total mass of stars in the in the inner 0.35 kpc and 1 kpc of the secondary, respectively; $M^{\rm sec}_{\rm y,\,{}0.35}$ and  $M^{\rm sec}_{\rm y,\,{}1}$: total mass of young stars ($<200$ Myr) in the inner 0.35 kpc and 1  kpc of the secondary, respectively.}
\end{center}
\end{table*}
%%%%%%%%%%%%%%%%%%%%%%%%%%%%%%%%%%%%%%%%%%%%%%%%%%%%%%%%%%%%%%%%%%%%%%%%%%%%%

The SFR in the secondary galaxy reaches a few $\sim{}10^{-3}$ M$_\odot{}$ yr$^{-1}$ at its maximum (slightly after the first pericentre passage), whereas the SFR in the bulge of the primary can be as high as $\sim{}5\times{}10^{-2}$ M$_\odot{}$ yr$^{-1}$ (Table~3), consistent with the {\it Swift}/UVOT observations of ESO 243-49.  The main difference between the two runs consists in the fact that the SFR in the centre of the secondary is quenched more gently with time in run~A (in which the velocity between the two CMs is higher, and there are just 2 pericentre passages in 4 Gyr), whereas it drops quite abruptly at $t\sim{}1$ Gyr after the first approach in run~B (in which the velocity between the two CMs is lower, and there are 5 pericentre passages in 4 Gyr).

Thus, the SF history derived from our simulations suggests that the ESO 243-49 system is in the late stages of a merger (after the second pericentre passage of the secondary), in which most of the SF takes place in the bulge of the S0. On the other hand, if  part of the blue and UV emission from the counterpart of HLX-1 is due to young stars (rather than to the reprocessing in the accretion disc), the SF associated with the secondary must not be completely switched off. Therefore, the  ESO 243-49 system might be in the stage between the first and the second pericentre passage, when the secondary still retains a sufficient SFR.

 A scenario where the nucleus of the secondary still hosts a (relatively) young stellar population is preferred to explain some properties of HLX-1. In fact, the X-ray variability of HLX-1 (with a possible periodicity of 380 days, Godet et al. 2009, Servillat et al. 2011) may be connected with a modulated mass transfer due to tidal stripping of a star in an eccentric orbit around the massive BH (L11). For this scenario to reproduce the observed properties of HLX-1, the mass of the companion star cannot be too small ($\approx{}3$ M$_\odot{}$, L11), although we stress that this requirement for the donor mass is the result of many assumptions (about the orbit of the star, the mass of the IMBH and the mechanism of accretion).
% many assumptions (about the orbit of the star, the mass of the IMBH and the mechanism of accretion) are underneath this requirement for the donor mass.}

\subsection{Evolution of the stellar and gas mass bound to the secondary galaxy}
 We now focus on the stellar and gas mass that remains bound to the secondary galaxy at a given time (Table~4). 
$M^{\rm sec}_{\rm \ast{},\,{}0.35}$ and  $M^{\rm sec}_{\rm \ast{},\,{}1}$ are the total mass of stars within 0.35 kpc and within 1 kpc from the centre of the secondary galaxy, respectively, at a given time (Table~4). The mass within 1 kpc gives an estimate of the stars that remain bound to the secondary galaxy, as the tidal radius is of the order of 1 kpc. Initially, $M^{\rm sec}_{\rm \ast{},\,{}1}$ is of the order of a few $\times{}10^8$ M$_\odot{}$. This is quite high with respect to the mass of $\approx{}10^6-10^7$ M$_\odot{}$, inferred for the optical counterpart of HLX-1. However, we stress that the values listed in Table~4 refer to the total stellar mass still bound to the stripped galaxy rather than to the (much denser) NC that can be hosted inside the nucleus: a NC cannot be dynamically resolved in our simulations (which have a softening of 0.1 kpc, i.e. approximately $100$ times the characteristic size of a NC). 
%The total stellar  mass inside 0.1 kpc is approximately one tenth of the  total stellar mass inside 1 kpc, more consistent with the observations of the HLX-1 counterpart.

The stellar mass within 0.35 kpc is reported for comparison with the VLT observations by S12 (which were taken with a seeing $\approx{}0.7-0.8$ arcsec, corresponding to $\approx{}0.35$ kpc for the distance of ESO~243-49). We take the observations by S12 as a reference, because the counterpart of HLX-1 is $\approx{}1$ magnitude fainter in S12 than in the {\it HST} data by F12, indicating that the contribution of disc reprocessing is lower. We converted the simulated values of $M^{\rm sec}_{\rm \ast{},\,{}0.35}$ into an estimate of the apparent $R$, $V$ and $B$ band magnitudes in the Johnson system, by assuming that the luminosity distance of ESO~243-49 is 95 Mpc, by adopting a Chabrier IMF (Chabrier 2001), and by using the Padova GALaxies AnD Single StellaR PopulatIon ModELs4 (GALADRIEL; Girardi et al. 2000) for a nearly solar metallicity ($Z=0.019$) stellar population. We stress that this procedure applied to N-Body/SPH simulations cannot be very precise, as it is limited by the intrinsic resolution of SFR and stellar ages in the simulations. We find that, at $t=4$ Gyr after the first pericentre passage, the apparent $R$, $V$ and $B$ magnitudes associated to $M^{\rm sec}_{\rm \ast{},\,{}0.35}$ are $R=20\pm{}1$, $V=21\pm{}1$ and $B=22\pm{}1$ for run~A, and   $R=24\pm{}1$, $V=25\pm{}1$ and $B=26\pm{}1$ for run~B.
The corresponding values reported by S12 for the counterpart of HLX-1 are $R=24.71\pm{}0.40$, $V=24.79\pm{}0.34$ and $B=25.19\pm{}0.30$. Thus, run~B at $t\ge{}4$ Gyr after the first pericentre passage is consistent with the VLT observations of the HLX-1 counterpart. The strongest constraint comes from the $R$ band, as the population associated with the simulated secondary galaxy is predominantly old. We expect that part of the light in the $B$ and $V$ bands is contributed by disc reprocessing. Instead, the nucleus of the secondary galaxy in run~A is still too bright at $t=4$ Gyr to match the observed photometry of the HLX-1 counterpart. We conclude that either the orbit simulated in run~A is not consistent with the evolution of HLX-1, or the elapsed time since the first pericentre passage is longer than 4 Gyr.

Therefore, the comparison between the luminosity of the simulated galaxy and the observed VLT photometry gives important constraints for the minor-merger scenario. In particular, the properties of the HLX-1 counterpart can be reproduced only if (i) a time $t\ge{}4$ Gyr elapsed since the first pericentre passage, or (ii) the orbit of the secondary galaxy is at least as bound as the orbit adopted in run~B (because a more bound orbit implies a faster tidal stripping, as the secondary galaxy undergoes more pericentre passages in a shorter time), or (iii) the mass ratio between the primary and the secondary galaxy is initially $>20$ (but galaxies significantly smaller than our simulated secondary galaxy are unlikely to host IMBHs at their centre).

%The hypothesis (iii) is unlikely, as galaxies significantly smaller than our simulated secondary galaxy are unlikely to host IMBHs at their centre. 
%On the other hand, we stress that mass to luminosity conversions in N-body simulations are subject to a number of uncertainties and that the spatial resolution limits our knowledge of the stellar distribution  in the nucleus of the secondary galaxy. Even the subtraction of the background light from ESO~243-49 might be an issue for the observational limits (see, e.g., S12).

Furthermore, our simulations suggest that the counterpart of HLX-1 is surrounded by a diffuse stellar halo, as $M^{\rm sec}_{\rm \ast{},\,{}1}>M^{\rm sec}_{\rm \ast{},\,{}0.35}$ for most snapshots. Thus, it is very important to assess whether the counterpart of HLX-1 is a completely naked star cluster or is surrounded by such diffuse stellar halo. The existence of such halo is not immediately evident from either {\it HST} or VLT images (F12;S12). This may be an issue for our model. On the other hand, from our simulations we expect the surrounding halo to be two (or more) magnitudes arcsec$^{-2}$ fainter than the central NC, i.e. very difficult to disentangle from the background contribution of the S0 galaxy. 
%Ideally, the presence of a faint stellar halo surrounding HLX-1  
%%and the total mass of old stars surrounding HLX-1 
%can help discriminating between various models of galaxy interactions: for example, the absence of a significant diffuse stellar halo surrounding HLX-1 might indicate that the time elapsed since the beginning of the interaction is longer than 3 Gyr, or that the initial stellar mass of the satellite galaxy was smaller, or that the orbit was less energetic. On the other hand, 
We must be very careful in interpreting what occurs at the (barely resolved) very centre of the simulated secondary galaxy, as spurious numerical effects can affect density profiles and tidal stripping. 
%%%A wider grid of simulations (including resolution tests), and a comparison with the available photometry, will be done in a forthcoming paper, to better constrain the initial properties of the satellite galaxy.
In a forthcoming paper, we will investigate (through a wider grid of simulations) for which orbital parameters the merger scenario can reproduce the observed photometry of the HLX-1 counterpart.

The total mass of gas in the inner kpc of the secondary galaxy after the first pericentre passage is $\sim{}1-2\times{}10^7$ M$_\odot{}$ (whereas most of the gas is already stripped in the tidal streams) and decreases by a factor of $\gtrsim{}10$ in the next 4 Gyr (Table~4). Interestingly, the total mass of young stars bound to the secondary galaxy (defined as stellar particles younger than 200 Myr) is of the order of $\approx{}10^5$  M$_\odot{}$ in the first Gyr after the first pericentre passage and drops to $\approx{}10^4$ M$_\odot{}$ in the following Gyrs. We stress that most of the young stars bound to the secondary galaxy are located in the inner 0.35 kpc, being much more concentrated than the old stars (Table~4). A young stellar component of  $\lesssim{}10^4$ M$_\odot{}$ is in agreement with the recent observations published by S12.

%Finally, Table~4 shows the line-of-sight velocity ($v_{\rm LOS}$) of the secondary, assuming the projection that best matches the observed location of the counterpart of HLX-1. As defined before, the plus (minus) sign indicates that the secondary is receding (approaching) with respect to the observer, i.e. it is redshifted (blueshifted). The modulus of $v_{\rm LOS}$ is almost always consistent with the observations (that suggest a redshifting by $\approx{}170$ km s$^{-1}$ with a large uncertainty, Wiersema et al. 2010), as a consequence of the assumed initial relative velocity between the two galaxies (Table~1). Therefore, the simulated systems spend most of their time in a kinematical state consistent with the available observations of HLX-1: more accurate spectroscopic measurements of HLX-1 counterpart are required to put stronger constraints.

%: Time vs mass gas, mass stars, SFR tot and loc

\subsection{Comparison with GalMer simulations}
%In the previous Section, we presented the results of two high-resolution simulations of galaxy mergers. 
%These give us important hints for ESO 249-43, but we cannot derive any statistical results from them, as too much CPU time is required to produce a statistically significant sample of such high-resolution simulations.
The two high-resolution simulations described in the previous sections provide important hints for ESO 249-43, but
%Two simulations, although with high resolution, 
do not allow any statistically significant considerations. On the other hand, it is noticeable that two simulations with completely different orbital properties (Table~2) give similar results, indicating that the evolution we describe is quite common for interactions between gas-rich dwarfs and S0 galaxies. 
%A statistical sample of simulations with the same properties as runs A and B will be performed for a forthcoming paper.
 
Interestingly, the main features of our runs~A and B (i.e. timescales, SF evolution and gas dynamics) are in agreement with the relatively low-resolution simulations of the GalMer database (Di Matteo et al. 2007; Chilingarian et al. 2010, and references therein). In particular, if we select out of the GalMer database the 36 runs describing an interaction between a giant S0 galaxy and a gas-rich disc dwarf galaxy, we can make the following remarks.
%derive the following considerations. 
(i) More than half of the considered GalMer simulations  (especially those with a retrograde orbit and the highest specific energy) develop the formation of a dense gas ring  surrounding the  bulge of the S0 galaxy, very similar to the one showed in Fig.~\ref{fig:fig5}. (ii) The interaction triggers the SF, initially at the centre of the secondary galaxy and, at later epochs, even in the nucleus of the primary: the SFR reaches (on average) a value of $0.08-0.1$ M$_\odot{}$ yr$^{-1}$, higher than in our simulation. However, this discrepancy can be explained by the different resolution (mass resolution in our runs is a factor of $\gtrsim{}10$ higher), by the different SF recipes and especially by the fact that the total gas mass is a factor of $\sim{}4$ higher in the GalMer sample than in our simulations. (iii) The average time spent by the simulated systems in a configuration similar to ESO 243-49 (i.e. in a configuration where the nucleus of the secondary is at $\lesssim{}10$ kpc projected distance from the centre of the S0 galaxy and where SF occurs in the nucleus of the S0) is $\sim{}700$ Myr, with respect to an average total duration of the merger of $\sim{}3$ Gyr (that is the average duration of mergers in the considered GalMer simulations). Thus, we can estimate that a S0 galaxy interacting with a secondary galaxy is in a `ESO~243-49-like configuration' for a fraction $f_{\rm close}\approx{}0.20$ of the duration of the merger.

There are substantial differences between our simulations and the GalMer sample: the mass-ratio between the secondary galaxy and the S0 galaxy is a factor of 2 larger in the GalMer database with respect to our simulations, the baryon-to-DM ratio is  up to a factor of $\sim{}10$ higher than in our simulations, the orbits span a slightly different parameter space with respect to ours (in GalMer $E_s\ge{}0$, and  $L_s$ is always slightly higher than our values), the recipes for SF are quite different, and the mass resolution is a factor of $\gtrsim{}10$ lower. %However, we find reasonable agreement between our results and those by GalMer. 
Taking into account all these differences, the agreement between the main features of our simulations and those of the GalMer database is noticeable.

\subsection{A statistical estimate}
In this Section, we estimate the density of ULXs similar to HLX-1 (hereafter, we will call them HLX-like sources, for simplicity), under the hypothesis that they are associated with a minor merger, and that they have an outburst X-ray luminosity $>10^{41}$ erg s$^{-1}$ (i.e., the minimum X-ray luminosity for a source to be defined a HLX, see e.g. King \&{} Dehnen 2005). In particular, we assume that the host of a HLX-like source is a gas-rich low-mass galaxy, undergoing merger with a giant galaxy. The requests that the secondary galaxy is gas rich and  undergoing merger are motivated by the fact that a gas-rich merger is needed to trigger SF. %%% (as a HLX-like source is expected to have a relatively young and massive stellar companion, L11).
In addition, we assume that the secondary galaxy has initially a stellar bulge in the $10^8-10^9$ M$_\odot{}$ range, i.e. the mass range of bulges that might host a central BH with mass $10^3-10^5$ M$_\odot{}$ and a NC  (Graham 2012a, 2012b). 
%{\bf In fact, it is unlikely that BHs with mass $<<10^3$ M$_\odot{}$ can reach an outburst luminosity $>10^{41}$ erg s$^{-1}$ (e.g., King & Dehnen 2005). }

  We define $\rho_{\rm host}$ as the stellar mass density of bulges that host HLX-like sources. $\rho_{\rm host}$ can be estimated as

\begin{eqnarray}\label{eq:eq1}
\rho_{\rm host}=360\,{}{\rm M}_\odot{}\,{}{\rm Mpc}^{-3}\left(\frac{h}{0.71}\right)\,{}\left(\frac{f_{\rm close}}{0.2}\right)\,{}\left(\frac{f_{\rm BH,NC}}{1}\right)\,{}\left(\frac{f_{\rm gas}}{1}\right)\,{}\times{}\nonumber{}\\
\left(\frac{g}{0.03}\right)\,{}\left(\frac{\rho_{\rm b}}{2.2\times{}10^8\,{}{\rm M}_\odot{}\,{}{\rm Mpc}^{-3}}\right)\,{}\left(\frac{f_{\rm merg}}{0.05}\right)\,{}\left(\frac{f_{\rm MT}}{0.02}\right)\,{}\left(\frac{f_{\rm duty}}{0.4}\right),
\end{eqnarray}
where $h$ is the Hubble parameter ($h=0.71\pm{}0.025$ according to the seven-year Wilkinson Microwave Anisotropy Probe data, Larson et al. 2011), $f_{\rm close}$ is the fraction of the merger timescale during which the nucleus of the secondary galaxy is sufficiently close to the disc of the primary (see previous section). $\rho_{\rm b}$ is the mass density of bulges in the local Universe ($\rho_{\rm b}=2.2\times{}10^8\,{}h\,{}{\rm M}_\odot{}\,{}{\rm Mpc}^{-3}$, according to Driver et al. 2007). $g$ is the mass fraction of bulges in the $10^8-10^9$ M$_\odot{}$ range (i.e., those that might host central BHs in the $10^3-10^5$ M$_\odot{}$ range, Graham 2012a). We adopt $g=0.03$, according to the Schechter formalism (Driver et al. 2007; Li \&{} White 2009). $f_{\rm BH,NC}$ is the fraction of bulges in the $10^8-10^9$ M$_\odot{}$ range that host both the central massive BH and a NC. We adopt $f_{\rm BH,NC}=1$, although this represents likely an upper limit (see, e.g., van Wassenhove et al. 2010, but Graham \&{} Spitler 2009 indicate that $f_{\rm BH,NC}$ may be of this order of magnitude). Finally, $f_{\rm gas}$ is the fraction of galaxies with a bulge in the $10^8-10^9$ M$_\odot{}$ range that host a significant amount of gas ($\approx{}10^8$ M$_\odot{}$). We put $f_{\rm gas}=1$, as an upper limit, although this value is uncertain.
 $f_{\rm merg}$ is the minor merger fraction in the local Universe. We assume $f_{\rm merg}=0.05$, in agreement with available data and models, although this number is  uncertain (see, e.g., D'Onghia, Mapelli \&{} Moore 2008; Jogee et al. 2009, and references therein). 

$f_{\rm MT}$ is the fraction of time that a BH similar to the one powering HLX-1 can spend in mass transfer with a companion star. There are no estimates of  $f_{\rm MT}$ for the case of HLX-1, and thus we adopt $f_{\rm MT}=0.02$ from Blecha et al. (2006), which is the predicted time spent in mass transfer by a $100-500$ M$_\odot{}$ BH hosted in a young star cluster. The actual value of $f_{\rm MT}$ may be very different from the assumed one, if the companion star is being tidally disrupted by the BH (L11) and/or if the BH is in a different environment from a young star cluster.  $f_{\rm duty}$ is the duty cycle, i.e. the fraction of time that the BH spends in a HLX state (i.e., X-ray luminosity -assumed isotropic- higher than $10^{41}$ erg s$^{-1}$). 
 We assume $f_{\rm duty}=0.4$ as, since its first detection, HLX-1 was observed above  $10^{41}$ erg s$^{-1}$ for $\approx{}150$ days during its $\sim{}380$ day semi-regular period (Godet et al. 2009; Servillat et al. 2011). However, $f_{\rm duty}$ is very uncertain, as we do not know the actual mechanism that powers HLX-1 and the light curve of HLX-1 covers only two periods, which are not particularly regular.

Therefore, the expected number density of HLX-like sources is 
\begin{eqnarray}\label{eq:eq2}
n_{\rm HLX}\simeq{}10^{-6}\,{}{\rm Mpc}^{-3}\,{}\left(\frac{h}{0.71}\right)\left(\frac{\rho_{\rm host}}{360\,{}{\rm M}_\odot{}\,{}{\rm Mpc^{-3}}}\right)\times{}\nonumber{}\\\,{}\left(\frac{3.7\times{}10^8\,{}{\rm M}_\odot}{\langle{}m_{\rm bulge}\rangle{}}\right),
\end{eqnarray}
where $\langle{}m_{\rm bulge}\rangle{}$ is the average mass of bulges in the $10^8-10^9$ M$_\odot{}$ range, according to the Schechter formalism (Li \&{} White 2009).
 This estimate suggests that there are $\approx{}4$ HLX-like sources in a sphere with a 100-Mpc radius. Since we now observe one HLX-1, either we are missing a fraction of HLX-like systems, or (more likely) we are overestimating some of the parameters in equation~(\ref{eq:eq1}). Likely, HLX-like sources have a duty cycle $<0.4$ (e.g., because other sources are in a different accretion regime with respect to HLX-1) and/or $f_{\rm MT}<0.02$. Another possibility is that  $f_{\rm BH,NC}<1$ or that $f_{\rm gas}<1$. Therefore, we can take our result as an upper limit, to be refined through new data.
\section{Conclusions}
HLX-1, hosted in the S0 galaxy ESO 249-43, is the brightest ULX known so far and the strongest IMBH candidate (F09). The optical counterpart of HLX-1 is a massive compact star cluster ($\gtrsim{}10^6$ M$_\odot{}$, S10; F12; S12), whose age is uncertain: {\it HST} observations (F12) are consistent with both a very young ($\approx{}10$ Myr) and a very old ($>10$ Gyr) massive  star cluster. While the optical and UV variability of the counterpart seems to exclude a young star cluster with a mass higher than $\sim{}10^4$ M$_\odot{}$ (S12), theoretical models suggest that the companion of the BH in HLX-1 cannot be too old, challenging even the old-cluster scenario (L11).  However, we note that the available constraints on the mass of the donor strongly depend on a number of theoretical assumptions and that alternative accretion mechanisms still need to be investigated (e.g., instabilities in a radiation-pressure dominated disc, L11).

In this paper, we studied, through $N-$body/SPH simulations, the scenario where the S0 galaxy ESO~243-49 is undergoing (or just underwent) a minor merger with a gas-rich low-mass disc galaxy. 
%ESO 243-49 is in a crowded environment (the cluster Abell 2877), which enhances the probability of a recent merger. 
The simulations show that the observed UV emission and the corresponding SFR ($\sim{}0.03$ M$_\odot{}$ yr$^{-1}$) in the bulge of ESO 243-49 can be explained as a consequence of the interaction. From the comparison between the observed SFR in the bulge of ESO 243-49  and the simulated SF history, we suggest that, if the UV emission from the optical counterpart of HLX-1 is mostly/entirely due to the reprocessing of the accretion disc, the ESO~243-49 system is currently at a stage after the second pericentre passage of the companion galaxy. In alternative, if part of the UV emission from the optical counterpart of HLX-1 is caused by a recent SF episode, the  ESO 243-49  system might be in between the first and the second pericentre passage of the companion galaxy. %{\bf Our simulations indicate that even 3 Gyr after the first pericentre passage the NSC of the satellite is still surrounded by a halo of stars (inside $\sim{}1$ kpc) which have not yet been stripped away. This may be an issue, as the existence of a faint stellar halo surrounding the optical counterpart of HLX-1 is not supported by available data. New simulations will help discriminating between various scenarios.}

We propose that the counterpart of HLX-1 coincides with the NC of the disrupting secondary galaxy. Recent studies (Graham \&{} Spitler 2009; Graham 2012a) indicate that low-mass galaxies with a $10^8-10^9$ M$_\odot$ stellar bulge can host both a NC and a $10^3-10^5$ M$_\odot$ IMBH. The NC scenario explains many properties of HLX-1: NCs are often composed by multiple stellar populations, mostly old stars with a younger population superimposed.
%a multiple stellar population, with a bulk of old stars and a younger superimposed population. 
Under this hypothesis, the counterpart of HLX-1  consists mainly of very old ($>10$ Gyr) stars, but the stellar companion of the IMBH can be a younger star. 
This satisfies the requirement that the mass of the companion star of the IMBH be not too small ($<<3$ M$_\odot{}$) to reproduce the observed X-ray luminosity and variability (see L11).
To further check our scenario, we need new UV observations and a simultaneous monitoring of the X-ray and UV emission, to understand whether the UV emission from the HLX-1 counterpart is entirely due to disc reprocessing or is partially connected with a young stellar population. 

 We compare the extrapolated luminosity of the simulated secondary galaxy with the observed magnitude of the HLX-1 counterpart (from the VLT observations reported by S12). We find that a very late merger stage ($\gtrsim{}4$ Gyr from the first pericentre passage), or a relatively bound orbit (eccentricity $\lesssim{}0.997$), or a primary-to-secondary galaxy mass fraction $\gtrsim{}20$ are required to be consistent with the observed $R$ magnitude of the HLX-1 counterpart. On the other hand, this constraint comes from many assumptions about the mass-to-light conversion in our simulations and may be affected by their spatial resolution. In a forthcoming paper, we will make a more accurate comparison with the observed photometry, by considering a wider grid of simulations (including a few higher resolution runs).

%-Crowded environment (rich group) enhances interactions

 If our scenario is correct,  we expect the density of HLX-like sources to be $\approx{}10^{-6}$ sources Mpc$^{-3}$, corresponding to $\approx{}4$ sources in a sphere of 100-Mpc radius.

From our simulations, we find that the three-dimensional distance of HLX-1 from ESO 243-49 might be much larger (even by a factor of $\sim{}20$) than the observed projected distance, if the merger is still ongoing. In fact, the existing spectroscopic measurements do not exclude a shift by $\sim{}170$ km s$^{-1}$, along the line of sight, between ESO 243-49 and the HLX-1 counterpart. Therefore, new spectroscopic measurements are required, to establish with better accuracy the relative velocity between ESO 243-49 and the HLX-1 counterpart, and to improve the constraints on the kinematics of the merger. In addition, there might be a number of X-ray sources analogous to HLX-1 that have been missed or mis-classified because they are (in projection) farther out from the host galaxy.
Therefore, it will be crucial to search for sources analogues to HLX-1, to test our and other possible scenarios.

\section*{Acknowledgments}
We thank the referee, Roberto Soria, for his useful comments, which helped to improve the paper. We also thank the authors of gasoline (especially J. Wadsley, T. Quinn and J. Stadel), L.~Widrow for providing us the code to generate the initial conditions, and L.~Girardi for making the results of GALADRIEL publicly available. To analyze simulation outputs, we made use of the software TIPSY\footnote{\tt http://www-hpcc.astro.washington.edu/tools/tipsy/\\tipsy.html}. We thank A. W. Graham, E. Ripamonti, R. Rampazzo and A. Marino for useful discussions. 
The simulations were performed with the {\it lagrange} cluster at the Consorzio Interuniversitario Lombardo per L'Elaborazione Automatica (CILEA).
We finally thank the GalMer team (I. Chilingarian, F. Combes, P. Di Matteo, A.-L. Melchior and B. Semelin) for making their simulation database publicly available\footnote{\tt http://galmer.obspm.fr/}.

%\onecolumn
%\appendix

\end{document}